\documentclass[runningheads]{llncs}

\usepackage{graphicx}
\usepackage{algorithm} 
\usepackage[misc]{ifsym}
\usepackage{algorithmic}

\usepackage{amsmath}
\allowdisplaybreaks
\usepackage{amssymb}
\usepackage{comment}

\usepackage{epsfig}
\usepackage{psfrag}

\usepackage{mathtools}

\usepackage{float}

\usepackage{tabularx}

\usepackage{algorithm}
\usepackage{algorithmic}

\begin{document}

 
\def\argmax{\mathop{\rm arg\,max}}

\title{A Consensus-Based Load-Balancing Algorithm for Sharded Blockchains}
%
\titlerunning{Consensus-Based Load-Balancing Algorithm for Sharded Blockchains}
%
\author{M. Toulouse\inst{1}$^{\textrm{\Letter}}$ \and H. K. Dai\inst{2} \and Q. L. Nguyen\inst{1}}
\authorrunning{M. Toulouse, H.K. Dai and Q. L. Nguyen}
%
\institute{School of Information and Communication Technology, Hanoi University of Science and Technology,
Hanoi, Vietnam\\
\email{michel.toulouse@soict.hust.edu.vn,lam.ntq166334@sis.hust.edu.vn}
\and 
Computer Science Department, Oklahoma State University,\\ Stillwater, Oklahoma 74078, U. S. A.\\
  \email{dai@cs.okstate.edu}} 
%
\maketitle              
\begin{abstract}
   Public blockchains are  decentralized networks where each participating node executes the same decision-making process. This form of decentralization does not scale well because the same data are stored on each network node, and because all nodes must validate each transaction prior to their confirmation. One solution approach decomposes the nodes of a blockchain network into subsets called ``shards", each shard processing and storing disjoint sets of transactions in parallel. To fully benefit from the parallelism of sharded blockchains, the processing load of shards must be evenly distributed. However, the problem of computing balanced workloads is theoretically hard and further complicated in practice as transaction processing times are unknown prior to be assigned to shards. In this paper we introduce a dynamic workload-balancing algorithm where the  allocation strategy of transactions to shards is periodically adapted based on the recent workload history of shards. Our algorithm is an adaptation to sharded blockchains of a  consensus-based load-balancing algorithm. It is a fully distributed algorithm inline with network based applications such as blockchains. Some preliminary results are reported based on simulations that shard transactions of three well-known blockchain platforms.

\end{abstract}
%
%
\section{Introduction}\label{intro}

 Replication in computer systems improves fault tolerance to network failures, computer hardware components failures, software bugs, malicious activities or to reduce systems access latencies. Databases have developed a variety of replication techniques as these systems face most of the above issues. Blockchain technologies extend the reach of replication to implement transparent decentralized control. 

The extensive use of replication in blockchains raises a whole set of familiar problematic issues which are often cited as a barrier to the adoption of this new technology. Cryptocurrency platforms such as Bitcoin and Ethereum can only process a limited number of transactions per second given that transaction validation is replicated across all the nodes of the blockchain network. Similarly, storage requirement grows proportionally to the \# of transactions $\times$ the number of nodes as the blockhain is identically copied on each node. Several solutions are proposed to address these two particular issues. One of them borrows from database systems, it is called sharding.

Sharding is a parallelization strategy for databases. Sharding partitions a large data set into multiple databases, where each database runs on a different server. Access requests to the data set are directed to the database where the requested data are stored.  As the databases run on different servers, requests can be served in parallel.   Designers of sharded blockhains  aim to exploit sharding parallelism to reduce the amount of memory needed to store blockchains and to speedup transactions processing.

Although sharding of blockchains has only been proposed recently, there are already several sharded blockchain protocols or fully implemented sharded blockchain platforms. Elastico \cite{2978389}, 2016, has been historically the first proposed sharding protocol. It has been followed by several others, among them  Zillica \cite{zilliqa-1}, Omniledger \cite{255586}, Chainspace \cite{chainspace}, RapidChain \cite{3243853}, Ethereum 2.0 \cite{EthereumShard}, Monoxide \cite{227661}, Ostraka \cite{Ostraka}, Harmony \cite{Harmony}, Logos \cite{Logos}, SSChain \cite{SSChain}, Stegos \cite{Stegos}, OptChain \cite{Nguyen_2019}, Sharper \cite{sharper}. 
The most widely known in the blockchain community   is probably Ethereum 2.0, a sharded version of Ethereum, to be released in 2021.  The following three recent surveys \cite{liu2021building,3355457,8954616} among others are summarizing design strategies related to sharded blockchains. 
 
The full sharding of a blockhain partitions the blockchain network into a set of sub-networks, i.e. shards, in which the nodes of each shard evaluates and stores disjoint subsets of the transactions generated by users. Known issues about this approach are cross-shard transactions and improper workload-balance among shards. Cross-shard transactions connect accounts which are stored in different shards, the evaluation of these transactions requires some form of synchronization, such as atomic commit protocols, among the shards involved in the cross-shard transactions.  Load-balancing for sharded blockhains consists to assign transactions to shards such that the sum of the transaction processing times is distributed evenly across shards. Mathematically, load-balancing is an NP-Hard optimization problem, it is unlikely polynomial time algorithms can be found to solve this problem. For sharded blockchains, load-balancing has an extra layer of complication as the processing time of transactions is unknown prior to assigning transactions to shards.

We are only aware of three recent publications addressing the issue of load-balancing in sharded blockchains \cite{krol2021shard,2020:1466,WooSK0P20}. All these papers propose centralized algorithms to periodically balance the workload of each shard. In the present paper,    
given that control decisions for blockchains are bottom-up, and achieved through consensus, we propose a fully distributed algorithm for sharded blockchain load-balancing problem, inline with the decentralization of blockchain designs. 
Our solution is based on distributed average consensus, a class of distributed algorithms to compute a global average of initial parameters using only local interactions among processing nodes. Distributed average consensus has been used for problems in diverse fields, control theory \cite{XiaoBK07}, multi-agent systems \cite{1205192}, physics \cite{PhysRevLett}, distributed optimization \cite{1104412}  including load-balancing, \cite{CYBENKO89}, 1989, where it is called {\it diffusion algorithm}. The diffusion algorithm solves the dynamic load-balancing problem, which is how we formulate the load-balancing problem for sharded blockhains. The work of \cite{CYBENKO89} has been extended to several dynamic load-balancing problems, parallel processing \cite{674316}, cloud computing \cite{doyle2012load}, many others. Our algorithm is based on one of these extensions \cite{674316}, adapted to the load-balancing problem for sharded blockchains from dynamic load-balancing techniques in
distributed processing.

Lastly, we have adapted two well-known scheduling algorithms for solving the shard load-balancing problem. The first one is the ``longest processing time" (LPT) algorithm from \cite{6767827} which assigns tasks to multiprocessors in decreasing order of their processing time. LPT minimizes the finishing time of the last executed task. Mathematically it is an approximation algorithm, which means it is proved the solutions return by LPT are no worst than some constant factor of the optimal solution.  Our implementation of LPT is centralized. The second algorithm is the MULTIFIT algorithm \cite{CoffmanGJ78}, also a scheduling algorithm minimizing the makespan of the schedule. MULTIFIT is also an approximation algorithm, compared to LPT it has a better approximation factor but its asymptotic running time is slightly worst than LPT. Our implementation of MULTIFIT is also centralized. We use the approximation factor of these two algorithms to validate the performance of the consensus-based algorithm. 

The paper is organized as follow. Next section briefly provides relevant background about blockchains, introduces a classification of sharding techniques for blockchains and  finally describes the workload problem specifically addressed in this paper. Section \ref{algo} describes three algorithms for our workload problem. Section \ref{s4} reports preliminary results of sharding simulations for three well known blockchain platforms. Finally, last section summarizes the paper and provides avenues for future investigations.

 \section{Problem Definition}\label{s2}

A blockchain can be described as a network of computing nodes running the client part of a blockhain application. The main purpose of a blockchain is to record the transfer of assets among  nodes (clients) of the network. A transfer of assets is manifested by the creation of a file called ``transaction" which has an id (the digest of a cryptographic hash function), and which holds data describing the transfer itself and  potentially some metadata for managing the transaction. 

The main processing activity of blockchain network nodes is to validate transactions,  as each transaction, prior to its confirmation, must be independently verified by all the nodes in the blockchain. Details of the verification process vary across blockchain platforms. Minimally, each node verifies whether the client node initiating the transaction holds the assets it transfers. The independent transaction verification is based on nodes having access on their local disk to the file of each previously posted transaction across the whole blockchain network. Once a transaction is confirmed,  the transaction file is indexed in a data structure called ``block". A block can hold a finite number of transactions, once this limit is reached, the block is closed and broadcast to the blockchain network to be stored on local hard disk of each node. Each block holds (in its header) the cryptographic hash of the previously stored block. Because of this cryptographic hash, the blocks form a chain, the ``blockchain". The blockchain must have the same sequence of blocks at every node, otherwise nodes will not be able to verify consistently ownership of the assets written for transfer in newly initiated transactions. All the transactions stored in the blockchain are visible to any user having an account on the blockchain.

\subsection{Types of Blockchains}
There are several criteria along which one can classify blockchain platforms. Two significant ones in regard of sharding protocols is the difference between public versus private blockchains and how the ownership of assets is recorded in the blockchain. 

\subsubsection{Public vs Private Blockchains}
Blockchain platforms can be divided into two categories: public, permissionless versus private, permissioned. In permissionless blockchains, an user only need to install the client side of the blockchain platform to attach  itself to the blockchain network,  without any form of authentication. Like credit cards, permissionless blockchains support applications targeting the consumer layer of the economy, they usually have a large network of nodes, Bitcoin and Ethereum, the two largest blockchain networks, are permissionless blockchains. Private blockchains support applications at the corporate level, which may demand a great level of trust among users, a high level of access securities, and which may place restrictions on which user may access which data in the blockchain. Private, permissioned, blockchain networks restrict entries, they have far fewer nodes compared to permissionless blockchains. In general, permissioned blockchains do not face the same pressing scalability issues as for public blockchains, they have much less transactions to process and to store. As a consequence, there are few private sharded blockchain proposals. In the list of sharded blockchains in Section \ref{intro}, only Chainspace \cite{chainspace} and Sharper \cite{sharper} are permissioned platforms.

\subsubsection{Transaction Models}
A second distinction among blockchains that impacts sharding designs is how assets are recorded.
 Most blockchain adopts one of two approaches: the Unspent Transaction Output (UTXO) model or the account-based model, the blockchains adopting one of these two models are said to be UTXO-based or account-based blockchains. In UTXO blockchains, assets are recorded only in the transactions. In this model, transactions have an input and an output section. Once a new transaction $tx$ is created to transfer assets, the input section of $tx$ refers to the output section of existing transactions as the source of assets to transfer. The output section of $tx$ list the public keys part of private keys that can unlock the assets transferred by $tx$. It is like someone write you a check, you give this check to pay for a given service, if your check is larger then the value of the service, you receive a new check for the difference in values. Your assets sums up to the checks you hold in your hand. In terms of UTXO blockhains, your assets are openly display on the blockchain, except it is not possible (in principle) to link the displayed assets to any specific user. 

The account-based transaction model is more intuitive. Users have anonymous accounts which are made up of a cryptographic pair of keys: public and private. Like bank accounts, each account has a balance. The account balance represents the value of the assets held by the owner of an account. Each transaction transfers assets from a source account to a destination account, the assets transfer is reflected in the balance of the accounts listed in the transaction. 

Account-based blockchains often have two types of accounts: user accounts and smart contract accounts. Smart contract accounts are computer programs stored on the blockchain. User accounts and smart contract accounts are stored on every node of the blockchain network.  When a transaction involves as source or destination a smart contract, the verification process of such transaction requires the smart contract be executed by each validating node. Similarly, the balance of user accounts listed in a transaction is verified by each node. The verification of transactions could not be executed independently if the user accounts or the smart contract accounts were not stored on each of the blockchain network node.

 In this paper, we address the load-balancing problem for public, permissionless account-based sharded blockchains. 

\subsection{Sharding Strategies}\label{s1}

Sharding strategies can be classified into 3 groups \cite{QingGSLZZ20}: state sharding, transaction sharding and full sharding.  State sharding means that the storage capacity collectively available from all the client nodes in the blockchain network is partitioned into shards, where each shard stores a disjoint subset of confirmed transactions. In state sharding, each transaction is verified by each node of the blockchain network. We are not aware of any state sharding protocol alone, but state sharding is often combined with transaction sharding in full sharding protocols.

In transaction sharding, computing nodes of the blockchain network are partitioned into subsets, called committees, where each committee verifies a disjoint subset of transactions. Transactions that are assigned to different committees are evaluated in parallel. Unlike state sharding, confirmed transactions are stored on all the nodes of the blockchain network, the storage requirement in transition sharding is the same as for un-sharded blockchains.
Zilliqa \cite{zilliqa-1} and Elastico \cite{2978389} implement this form of sharding.  Zilliqa is an account-based blockchain. Computing nodes are partitioned into a pre-defined number of committees. Transactions are assigned to committees for processing based on the address of the account that initiates the transaction.
For example, if there are $l$ committees from 0 to $l-1$, the last $\lfloor \log_2 l \rfloor + 1$ bits of the initiating account address, a number between 0 and $l-1$, identifies the committee where the transaction is sent for processing. 
 Elastico, an UTXO transaction model, assigns transactions to committees according to the transaction hash id. For example, if there are 8 committees, the decimal conversion of the first 3 bits of a transaction id defines the committee assignment of the transaction.

In full sharding, storage and transaction verifications are sharded. Several recent blockchain platforms/protocols apply this sharding design: RapidChain (UTXO) \cite{3243853}, SSChain \cite{SSChain} (UTXO), Omniledger \cite{255586} (UTXO), Harmony \cite{Harmony} (account-based), Monoxide \cite{227661} (account-based) and  Ethereum 2.0 (account-based). Computing nodes are assigned to committees according to rules that vary across these sharded platforms. In full sharding for account-based blockchains,  the storage of account balances as well as smart contract codes is also sharded, that is user accounts and smart contracts are stored only on the computing nodes of the committee where they are assigned. Consequently, the assignment of transactions to shards is implicitly done through  accounts, i.e. transactions are verified in the committee where the account initiating the transaction is stored.   
In full sharding, each shard is responsible for generating blocks from transactions confirmed in the shard, for maintaining its own blockchain and for storing a subset of the account balances. Full sharding divides the storage requirement of a blockchain platform by the number of shards. Note that in order to defend against security vulnerabilities, some or all the nodes forming a committee are re-assigned periodically to other shards. 

\subsection{The Load-Balancing Problem for Sharded Account-Based Blockchains}\label{lb}

The general form of the load-balancing problem is formulated as follow: assign $m$ tasks $t_1, t_2, \ldots, t_m$ with respectively computing times $c_1, c_2, \ldots, c_m$, to $n$ processing units $p_1, p_2, \ldots, p_n$  such to minimize the maximum load on any processing unit. Assume that ${\cal T}_i$ is the set of tasks assigned to processing unit $p_i$, the load on machine $p_i$ 
is $C_i = \sum_{k \in {\cal T}_i} c_k$, the objective function is $\min \max\{C_i \mid i = 1..n\}$.

The load-balancing problem for account-based sharded blockchains can be modeled as follow: given a set of $m$ accounts $a_1, a_2, a_m$, assign the $m$ accounts to $n$ shards $s_1, s_2, \ldots, s_n$ such to minimize the maximum load on any shard. Let ${\cal R}_j$ be the set of transactions initiated by account $a_j$ and $p^j_k$ the processing time of transaction $k$ initiated by $a_j$. The  computing time $c_j$ of account $a_j$ is the sum of the processing times $p^j_k$ of the transactions initiated by account $a_j$, $c_i = \sum_{k \in {\cal R}_j}p^j_k$. Let ${\cal T}_i$ be the set of accounts assigned to  shard $s_i$, the workload of shard $s_i$ 
is $C_i = \sum_{k \in {\cal T}_i} c_k$, the objective function is $\min \max\{C_i \mid i = 1..n\}$.

The sharded load-balancing problem is not static, new accounts are constantly created from existing users or new users joining the blockchain network, and transactions are constantly initiated from the account holders. Thus the load of shards is changing, it has to be continuously re-evaluated, which is the definition of a dynamic load-balancing problem. The computing time of shards is divided in periods between which load-balancing is performed. The duration of the periods could be variable based on measurements of the shards load imbalances. In this paper, periods have fixed elapse times, we name these periods ``epochs".

The processing time of an account is not known prior to assign an account to a shard, thus load prediction must be applied. Consider two consecutive epochs $e_k$ and $e_{k+1}$. During epoch $e_k$, the real processing time of each account $a_i$ is recorded in $c_i$. The processing time $c_i$ from epoch $e_{k}$ is used as a prediction of the processing time of account $a_i$ in epoch $e_{k+1}$. Load-balancing is performed before epoch $e_{k+1}$ starts, the balanced workload for $e_{k+1}$ is computed based on predictions from epoch $e_k$. 

\section{Algorithms}\label{algo}

This section describes three algorithms for solving the dynamic load-balancing problem formulated in the previous section: the diffusion algorithm, LPT and MULTIFIT.

\subsection{Diffusion Algorithm}
 
The diffusion algorithm belong to a class of algorithms called distributed average consensus algorithms which compute global averages in parallel in decentralized networks such as ad-hoc networks, peer-to-peer networks or sensor networks. Mathematically the network is represented as an undirected graph where adjacency among the graph vertices stands for direct communication links among the network computing nodes.  Let $G = (V,E)$ be such a graph where $V = \{v_1, v_2, \ldots, v_n\}$ is a set of vertices (modeling the network computing nodes) and $E$ denotes a set of edges pairing vertices (direct communication links). Graphs like $G$ have an adjacency structure  represented by some $n \times n$ {\it adjacency matrix} (denoted by $A$ here) where
 $a_{ij} = 1$ if and only if $(v_i, v_j) \in E$, $a_{ij} = 0$ otherwise. The adjacency structure of $G$  defines for each node $v_i \in G$ a {\it neighborhood} ${\cal N}_i$ where ${\cal N}_i = \{v_j \in V | (v_i, v_j) \in E\}$. The global average computed by distributed consensus is $\frac{1}{n} (\sum_{i=1}^nx_i(0))$  where $x_i(0)$ is some initial value of vertex $i$. The global average is obtained from the execution by each node of the network of the following iterative algorithm: 
\begin{equation}
x_i(t+1) = w_{ii}x_i(t) + \sum_{j \in {\cal N}_i} w_{ij}x_j(t)
\label{eq1}
\end{equation}
where $w_{ij}$ is a weight associated to edge $(i,j)$. The weight values $w_{ij}$ belong to a $n \times n$ weight matrix $W$ satisfying  some properties such as $W = W^T$ (transpose of $W$) and others algebraic properties that are used to prove the local updates converge asymptotically to the global average $\frac{1}{n} (\sum_{i=1}^nx_i(0))$. The Metropolis-Hasting weight matrix below satisfies those conditions \cite{XiaoBK07}:

\begin{eqnarray} W_{ij} = \left\{ \begin{array}{ll}
 \frac{1}{1 +\max(d_i,d_j)} &\hspace{2mm}\mbox{if $i \not = j$ and $j \in {\cal N}_i$} \\
  1 - \sum_{k \in {\cal N}_i} W_{ik} & \hspace{2mm}\mbox{if $ i = j$}\\
  0 & \hspace{2mm}\mbox{if $i \not = j$ and $j \not \in {\cal N}_i$}
       \end{array} \right. \label{aa}\end{eqnarray}
where $d_i = |{\cal N}_i|$.

In order to solve the shard load-balancing problem using a consensus algorithm, shards must be embedded in some network topology defining the neighborhood of each shard. For simplicity, we assume shards are connected through a ring network, thus a regular graph, where each shard has two neighbors. The weight matrix is Metropolis-Hasting matrix, wherein a ring network, $w_{ij} = 0.33$ if $a_{ij} = 1$, $w_{ij} = 0$ otherwise. Shard processing is synchronous, divided into fix length epochs. The initial value $x_i(0)$ in the consensus model is the sum of the processing times of shard $i$, its workload.
Each iteration of Equation (\ref{eq1}) computes the local workload average of shard $i$ and its neighbors. This procedure converges locally to the global average of the shards workload. 

The diffusion algorithm  computes local load averages and the amount of load to transfer between neighbor shards once the distributed average consensus algorithm has converged to a workload balance among all shards. We use the same distributed average consensus as in \cite{CYBENKO89}, it differs slightly from Equation~(\ref{eq1}):

\begin{equation}
Load_i(t+1) = Load_i(t) - \sum_{j \in {\cal N}_i} w_{ij}(Load_i(t) - Load_j(t))
\label{eq3}
\end{equation}
where $Load_i(t)$ is the load of shard $i$ at iteration $t$. The full description of the diffusion algorithm appears in the pseudo-code of {\bf Algorithm \ref{phase1}}. This algorithm is executed by each shard. 

\begin{algorithm}[h]
\caption{Diffusion algorithm for sharded blockchain load-balancing}\label{phase1}
\begin{tabbing}
aa\=aa\=aa\=aa\=aa\=aa\=aa\=aa\=aa\=aa\=aa\kill\\
Diffusion$\_$algorithm($i$, ${\cal N}_i$, $n$, $W$, workload of shard $i$)\\
\> int $Load_i(0) =$ workload of shard $i$\\
\> int $t = 0$\\
\> float $\Delta_i[n]$ = 0\\
\>while no convergence\\
\>\>for ($j=0; j < n, j++$)\\
\>\>\> if $j \in {\cal N}_i$\\
\>\>\>\> $\Delta_i[j](t+1) = \Delta_i[j](t) + w_{ij} (Load_i(t) - Load_j(t))$\\
\>\>$Load_{i}(t+1) = Load_{i}(t)-\sum_{j \in {\cal N}_i} w_{ij}(Load_{i}(t) - Load_{j}(t))$\\
\>\>$t = t +1$
\end{tabbing}
\end{algorithm}

$Load_i(0)$ is the load of shard $i$ at the end of an epoch (equivalent of $x_i(0)$ in the consensus model). $\Delta_i$ is a vector of $n$ entries, called {\it transfer vector} in \cite{674316}. At each iteration $t$ of the ``while'' loop, $\Delta_i[j](t+1)$ stores the load that shard $i$ must transfer to its neighbor $j$ such that the load of shard $i$ at iteration $t+1$ is the local average of its load and its neighbor loads at iteration $t$, i.e. $Load_i(t+1) = \frac{Load_i(t) + \sum_{j \in {\cal N}_i} Load_j(t)}{|{\cal N}_i| + 1}$, note $\Delta_i[j] = 0$ if $j \not \in {\cal N}_i$.

Computationally, instruction $Load_{i}(t+1) = Load_{i}(t)-\sum_{j \in {\cal N}_i} w_{ij}(Load_{i}(t) - Load_{j}(t))$ is not necessary as the load of a shard at iteration $t+1$ can be computed using the transfer vectors and the loads of the previous iteration. However this instruction allows us to claim without proof that Algorithm \ref{phase1} converges asymptotically to $\frac{\sum_{i=1}^{n} Load_{i}(0)}{n}$ based on numerous convergence proofs in the literature for iterative procedures such as Equations (\ref{eq1}) and (\ref{eq3}).  

Algorithm \ref{phase1} converges to a state where the local averages are all approximately the same, the same as the global average, and where the loads are balanced among all shards. 
This is however not yet the case after the first iteration of the algorithm. At iteration 1 of the ``while'' loop, the load of shard $i$, $Load_i(1)$,  is the average load of shard $i$ and its neighbors at iteration 0, i.e. $Load_i(1) =  \frac{Load_i(0) + \sum_{j \in {\cal N}_i} Load_j(0)}{|{\cal N}_i| + 1}$. However the load of shard $i$ is not the same as the average load of shard $i$ and its neighbors at iteration 1: \linebreak $Load_i(1) \not =  \frac{(\sum_{j \in {\cal N}_i} load_j(1)) + load_i(1)}{|{\cal N}_i| + 1}$, the loads are not balanced at iteration 1. Average loads is not equivalent to balanced loads initially. It is only once the average loads are all the same across all shards, where they are the same as the global average, that we have the workloads to be balanced.

For all the iterations of Algorithm \ref{phase1}, the transfer vector $\Delta_i[j]$ always contents the load to transfer from shard $i$ to shard $j$ to get local average loads among neighbor shards. However, once the local averages are the same as the global average, where the global average is the balanced load, the values in the transfer vectors take another meaning, they represent the amount of load to transfer among neighbor shards to get a balanced workload, they are the solution to the load balancing problem. 

At the conclusion of Algorithm \ref{phase1}, if $\Delta_i[j]$ is positive, shard $i$ must send to shard $j$ some load. If negative, shard $i$ must received some load from shard $j$. 
The behavior of Algorithm \ref{phase1} is illustrated in Table \ref{table 6} using an extreme case of initial load imbalance. It consists of an instance of 5 shards  0, 1, 2, 3,  4, with initial loads 0, 26, 0, 0, 0. 

\begin{center}
\begin{table}
{\scriptsize
\begin{tabular}{c|ccccc||cc|cc|cc|cc|cc|cc}
&\multicolumn{5}{c||}{Shards} & \multicolumn{10}{c|}{Transfer vectors}\\\hline
Iters&0&1&2&3&4 &$\Delta_0[4]$&$\Delta_0[1]$&$\Delta_1[0]$&$\Delta_1[2]$&$\Delta_2[1]$&$\Delta_2[3]$&$\Delta_3[2]$&$\Delta_3[4]$&$\Delta_4[3]$&$\Delta_4[0]$\\\hline
0&0&26&0&0&0&0.00&-8.58&8.58&8.58&-8.58&0&0&0&0&0\\
1&8.58&8.84&8.58&0&0&2.83&-8.66&8.66&8.66&-8.66&2.83&-2.83&0&0&-2.83\\
2&5.83&8.66&5.83&2.83&2.83&3.82&-9.60&9.60&9.60&-9.60&3.82&-3.82&0&0&-3.82\\
3&5.77&6.79&5.77&3.82&3.82&4.46&-9.93&9.93&9.93&-9.93&4.46&-4.46&0&0&-4.46\\
4&5.46&6.12&5.46&4.46&4.46&4.79&-10.26&10.26&10.26&-10.26&4.98&-4.98&0&0&-4.98\\
5&5.35&5.69&5.35&4.69&5.35&&&&&&\\
$\vdots$&&&&\\
8&&&&&&5.16&-10.37&10.37&10.37&-10.37&5.16&-5.16&0&0&-5.16\\
9&5.21&5.24&5.21&5.16&5.16&&&&&&&&&&\\
$\vdots$&&&&\\
14&&&&&& 5.19&-10.39&10.39&10.39&-10.39&5.19&-5.19&0&0&-5.19\\
15&5.20&5.20&5.20&5.19&5.19&&&&&&&&&\\
\end{tabular}}

\vspace{5mm}
\caption{Illustration of Algorithm \ref{phase1}} \label{table 6}
\end{table}\end{center}

In Table \ref{table 6}, given an ``Iters'' value $i$, the ``Shards'' columns list the workload of each shard while the ``Transfer vectors'' columns provide the values of the two relevant entries of each transfer vector, for example $\Delta_0[4]$ $\Delta_0[1]$ is the transfer vector of shard 0. At iteration 0 the algorithm computes the transfer vectors for the loads obtained from the previous epoch. For example, shard 0 receives no load from shard 4 but receives 8.58 load from shard 1 as shown in $\Delta_0[4]$ $\Delta_0[1]$. We can see also that at iteration 0 shard 1 send 8.58 load to shards 0 and 3. As a consequence of the transfers computed at iteration 0, the loads of shards 0 and 2 is 8.58 at iteration 1. We can also see that each workload at iteration 1 is the average of the workloads at iteration 0 (modulo some rounding errors) and that the average loads at iteration 1 differ among shards, thus the load among the shards is not balanced.

The diffusion process starts with the transfers computed at iteration 1 where shard 0 and 2 sends respectively 2.83 load to shard 4 and shard 3. This is diffusion of the load originally in shard 1 to shards which are not neighbors of shard 1. A similar diffusion process takes place with the transfer vectors. Eventually the average consensus algorithm converges to balanced loads, such as in iteration 15. At this point, as in iteration 14, transfer vectors store the value of the load that must be transferred from each shard to its neighbors such that each shard starts the next epoch with a balanced workload. 

Algorithm \ref{phase1} computes the load that must be transferred, but it  does not actually migrate the accounts that actualize these workload transfers.
The next step consists  to migrate accounts for which the sum of the processing times is equivalent to the transfer values computed by  Algorithm \ref{phase1}. 
During this step, each shard $i$ must find $|{\cal N}_i|$ subsets of accounts such that the sum of the processing times  of the accounts in a subset is equal to a positive $\Delta_i[j]$, $j \in {\cal N}_i$. Computing these subsets is equivalent to solve the subset sum problem, a problem that is known to be NP-Hard. We use a heuristic to compute these subsets. The accounts in a shard  $i$ are sorted in decreasing order of their processing time. Then the list is traversed from the largest processing time  to the smallest one, when the processing time of an account is smaller than a current positive $\Delta_i[j]$ value, the corresponding account is selected to be migrated to shard $j$. 

It may not be possible for all shards to complete the accounts migration in one round. The load of some shards could be smaller than the positive values it needs to transfer to its neighbors. In an extreme case, the workload might be concentrated in one shard $i$ (such as shard 1 in Table \ref{table 6}), thus the migration of accounts in the first round is only possible from shard $i$ to its neighbors. In such extreme load imbalance, the migration of accounts may only be completed after several rounds, where in each round  accounts migration only partially fulfill the vector transfers. However,  according to \cite{674316}, it can be shown that the theoretical number of rounds is bounded by the diameter of the network, which is consistent with a lower-bound result in \cite{Dai2020}, in practice we have observed that the number of rounds is much smaller.  

\subsection{Centralized Algorithms}

In this section we introduce two algorithms for the independent task scheduling problem on multiprocessors. The canonical formulations of this  problem and load-balancing are very similar. Algorithms for  load-balancing are used for solving the task scheduling and vice-versa.  Assuming accounts are independent, the load-balancing problem for sharded blockchain easily conforms to the independent task scheduling problem. 

The algorithms described in this section  are well-known approximation algorithms for the static independent task scheduling problem: the Longest Processing Time (LPT) and the MULTIFIT algorithm. The two algorithms are proposed as centralized solutions. They have been adapted to solve the dynamic load-balancing problem for sharded blockchains.

\subsubsection{Longest Processing Time}

Longest Processing Time  solves load-balancing for shards by first sorting the accounts in decreasing order of their respective processing times. In that order, the algorithm sequentially assigns accounts to the  shard with the smallest load so far. 

The easiest way to implement the sorting phase of this algorithm in a distributed computing environment is to send the processing time of each account to a single network node. This node sorts the processing times and migrates accounts to shard if their assignment  differs from the one they had in the previous epoch.  According to \cite{6767827}, LPT has an approximation factor of $(\frac{4}{3}- \frac{1}{3m})$ (about 1.58) of the optimal solution. In 
\cite{DellaCroce2020}, the authors report a new approximation factor of $(\frac{4}{3}- \frac{1}{3(m-1)})$.
The asymptotic running time of LPT is $O(n \log n+n\log m)$ ($n =$ number of shards,  $m =$ number of accounts).

\subsubsection{MULTIFIT}
MULTIFIT, first introduced in \cite{CoffmanGJ78}, is a second approximation algorithm for solving the independent task scheduling problem. MULTIFIT repetitively  solves a bin packing problem. A set of $n$ bins (shards) are assigned with a same capacity $C = \frac{1}{2}(A = \frac{\sum_{i=1}^m c_i}{n} + B = 2\frac{\sum_{i=1}^m c_i}{n})$ ($m =$ number of tasks, $c_i =  $ processing time of task $i$). The tasks are sorted in decreasing order of their respective processing times. In that order, the tasks are assigned into a first bin until a $c_i$ overload the capacity $C$ of the current bin. Then task $t_i$ is assigned to the next bin, this continue until all the tasks are assigned in the $n$ bins, or until all the $n$ bins are full. If all the tasks have been assigned then a new bin packing round starts with a lower capacity $C' = \frac{1}{2}(A + (B = C))$ for each bin. If not all the tasks have been assigned in the previous round, then the new round starts with an increased capacity $C' =\frac{1}{2}((A = C) + B)$. 
This bin packing problem is solved for $k$ rounds. The value of $B$ in the last round is the length of the longest schedule.  The asymptotic running time of MULTIFIT is $O(n \log n + kn \log m)$, and the approximation factor is 1.22 + $(\frac{1}{2})^k$. Compared to LPT, MULTIFIT has a slightly worst asymptotic running time but it has a tighter approximation factor. 

The repeat rounds of solving the bin packing problem can only be executed by a single node. Thus the implementation of MULTIFIT for the shard load-balancing problem follows a similar pattern as for the LPT algorithm. After the last round, bins are mapped to shards, the content (accounts) of each bin is migrated to the respective mapped shard, if needed. 

\section{Experimentation}\label{s4}

This section provides preliminary results from tests which simulate the sharding of three account-based blockchains: Ethereum, Zilliqa and Binance Smart Chain. Data for the simulations are obtained from crawling the following transactions tracker websites: Ethereum (\url{https://etherscan.io/}),  Zilliqa \linebreak (\url{https://viewblock.io/zilliqa})  and Binance Smart Chain (\url{https://bscscan.com}). Ethereum 2.0, a sharded version of Ethereum, is only known at the moment of writing this paper as protocol, thus here we use Ethereum transactions. The transactions tracker website for Zilliqa does not list with  which shard transactions and accounts are associated. Binance Smart chain is a non-sharded account-based blockchain.

\begin{table}[h]
\centering
\begin{tabular}{l | l | l | l }
     & Ethereum & Binance & Zilliqa \\
\hline
Number of transactions & 44511 & 49989 & 49980 \\
Number of accounts & 21548 & 5507 & 12487 \\
Number of shards & 10 & 10 & 10 \\
Accounts per shard (approximately) & 2155 & 551 & 1249 \\
\end{tabular}
\caption{Number of accounts and transactions per blockchains}
\label{tab:abc}
\end{table}

The data collected per transaction are the  following ones: transaction hash id, block
hash id where the transaction is indexed, the account source, the account destination, the time the transaction processing started, the amount of asset transferred in the transaction and the transaction fees. We have tracked 44511 Ethereum transactions, 49989
Binance transactions and 49980 Zilliqa transactions. Table \ref{tab:abc} lists the number of source accounts for the transactions that have been crawled.

\begin{figure}[th]
                \centering                \includegraphics[width=0.75\textwidth]{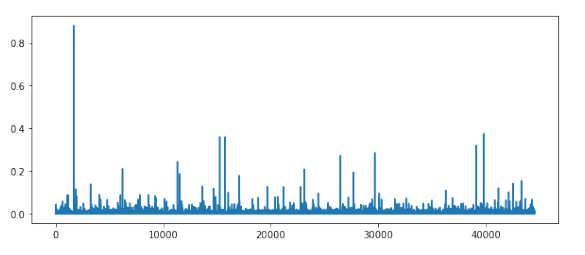}
                \caption{Transaction fees for the 44522 Ethereum transactions}\label{f1}
            \end{figure}

\subsection{Sharding Simulations}
The simulations only implement the most basic components of account-based shard designs such as the assignment of accounts and transactions to shards. 
The processing time of Ethereum transactions is reflected into the transaction fees expressed in ``gwei", one gwei =  0.000000001 ETH \cite{gasETH}, the native Ethereum currency. The fees that users must pay for processing a transaction are largely based on the number of CPU cycles spent to verify the transaction and some other marginal considerations such as the size of the transaction. The transaction fees expressed in ETH, as reported on the tracking website, is used here as a measure of transaction processing times. Processing times for Zilliqa and Binance transactions are also based on  transaction fees in a similar way.   

\begin{figure}[h]
    \centering
    \includegraphics[width=0.90\textwidth]{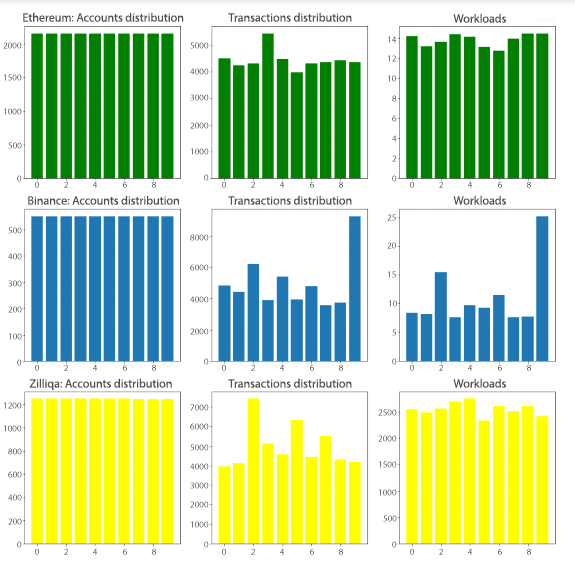}
    \caption{Shards workload and transaction distributions}\label{f2}
    \end{figure}

Prior to run our load-balancing algorithms,  we have excluded some outlier transactions from the crawled data. An outlier transaction is one for which the transaction fees are disproportionally high compared to the vast majority of the transactions. Figure \ref{f1} illustrates this filtering process for Ethereum transactions where those transactions that cost more than 0.2 ETH are excluded, 11 transactions have been filtered out. Outlier transactions obscure at the present stage the comparative analysis of the performance trends for the three load-balancing algorithms. The number of transactions listed in Table 
\ref{tab:abc} is after removing outliers.

We simulate 10 shards for each of the three blockchain platforms. In account-based blockchains, shards are constituted by assigning each account to a shard. The assignment of accounts to shards  can be done by users. Alternatively, accounts can be assigned by the sharded platform which statically assigns accounts to shards.  Our simulation follows the last approach. The accounts from the crawled transactions are assigned randomly and as evenly as possible to shards.  This initial distribution of accounts per shards is reported in the first column of charts in Figure \ref{f2}, in which the first, second and third rows stand respectively for results from Ethereum, Binance and Zilliqa. This figure shows that the number of accounts per shard is distributed uniformly across all shards and blockchain platforms. The second column of charts in Figure \ref{f2} reports the number of transactions per shard given  the initial random assignment of accounts to shards. The number of transactions per shard varies substantially for Binance and Zilliqa.

The simulation runs for 2 epochs. For the first epoch, the transactions fees of transactions originating from a same account are summed up together, this sum represents the processing time of the corresponding account. The sum of the processing time of the accounts assigned to a same shard is computed, which is the workload of the shard. The third column in Figure \ref{f2} reports the initial workload of each shard.  For Binance, the imbalance in the number of transactions for the last shard translates in an imbalance in the workload for this same shard.

\begin{figure}[th]
    \centering
    \includegraphics[width=0.75\textwidth]{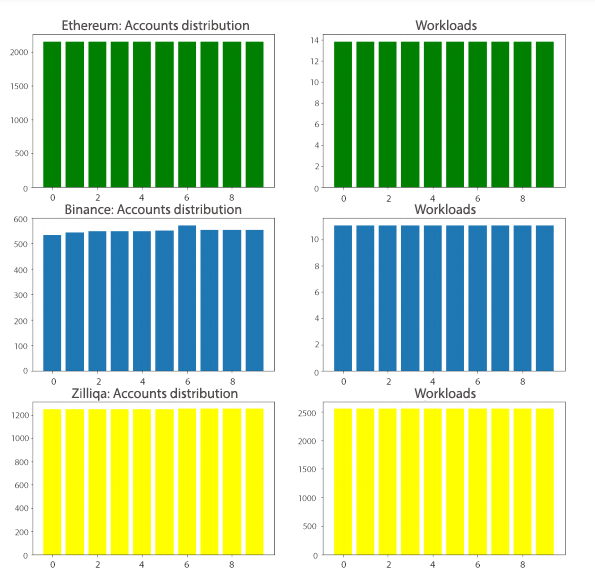}
    \caption{Shards workload-balance using LPT algorithm}\label{f3}
\end{figure}

\subsection{Numerical Results}

\begin{figure}[th]
    \centering
    \includegraphics[width=0.75\textwidth]{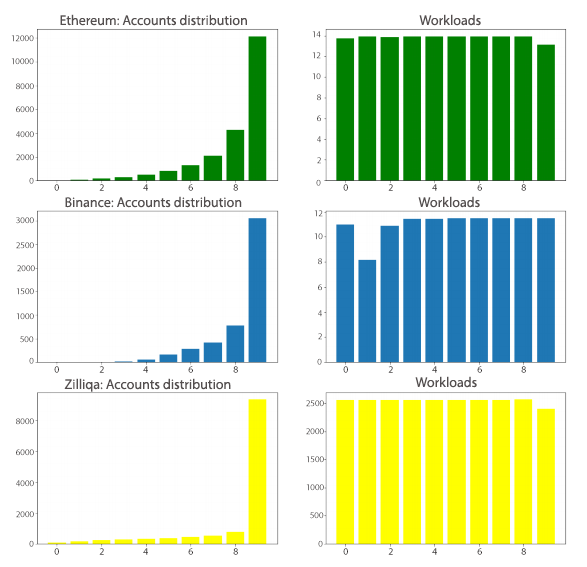}
    \caption{Shards workload-balance using MULTIFIT algorithm}\label{f4}
\end{figure}

\begin{figure}[th]
    \centering
    \includegraphics[width=0.75\textwidth]{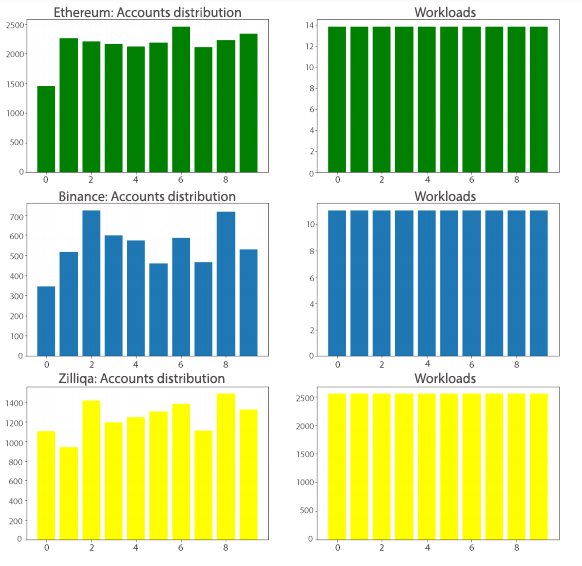}
    \caption{Shards workload-balance using the diffusion algorithm}\label{f5}
\end{figure}

		The three algorithms described in Section \ref{algo} are executed at the end of epoch 1 using as input the processing time of each account as calculated for epoch 1. According to the outputs of these algorithms, accounts are re-assigned to shards. Then the workload of each shard is computed in the same way as for epoch 1 using the same set of transactions (after re-assigning accounts). Figures \ref{f3}, \ref{f4} and \ref{f5} report the computed workloads of epoch 2 respectively for the LPT, MULTIFIT and the diffusion algorithm. The first column of charts in each of these three figures report the number of accounts per shard while the second column of charts report the workload per shard.

		The LPT algorithm in Figure \ref{f3} shows an excellent workload-balance per shard as well as quite even distributions of the accounts per shard for the three sharded blockchain simulations. For the MULTIFIT algorithm, Figure \ref{f4}, we can see that the workload of shard 10 for Ethereum and Zilliqa is off a little bit. However, the striking aspect of the MULTIFIT solutions is the irregular distribution of accounts per shard. MULTIFIT sorts accounts in decreasing order of their transaction fees. Thus the first accounts fill the bins (shards) with relatively few accounts while at the low end of the sorted accounts, many accounts are needed to fill a single bin. As a result, the number of accounts in the last shards is substantially larger than those in beginning shards. 
		
		Last, Figure \ref{f5}, reports the performance of the diffusion algorithm. The load-balance for the diffusion algorithm appears as good as LPT. This indicates that workloads are as good as the approximation factor of the LPT algorithm. Although not shown in the charts, the second phase of the diffusion algorithm was completed in one round for all the tests.
		
The number of tests reported in this section is too small to draw general conclusions about these algorithms. Further tests are required over larger sets of crawled transactions, over several epochs, each epoch using a different set of crawled transactions. Computing times of each load-balancing algorithm will have to be recorded and compared. Intuitively, we expect the diffusion algorithm to have better computing times as it is a parallel algorithm.  Outlier transactions would have to be included in the test sets to evaluate the robustness of the algorithms.

However, based 	on Ethereum online statistical data available at \linebreak \url{https://bitinfocharts.com/ethereum}, it is not expected  that significant divergences will exist between our preliminary results and results of more extensive simulations. Statistics about Ethereum data extracted from Table \ref{tab:abc} and Figure \ref{f2} conform with the daily reported statics on the above website. The number of initiated transactions per  account is about 2 in Table \ref{tab:abc}, similar to the daily average number of initiated transactions per 
account. Transaction fees in Figure \ref{f2}, which are on average about 0.003 ETH per transaction, are quite similar to the daily average  transaction fees reported for Ethereum. Simulations conducted with a large sample of Ethereum transactions are likely to have similar workload distributions as the distributions in our current tests. Unfortunately, similar online statistics are not available for Binance and Zilliqa. 

\subsection{Discussion}

Load-balancing adds vulnerabilities and computing/communication overheads to the  operation of sharded blockchains. The diffusion algorithm mitigates some of these issues, it is fast as it runs in parallel, it is not susceptible to single point of failure issues such as DoS attacks or other attacks aiming at controlling the load balancing protocol executed by a single node and migration of accounts is also performed in parallel and is local which likely cost less in terms of communication. Centralized algorithms such LPT and MULTIFIT rank low under these criteria, account processing times must be forwarded to a single server, account scheduling to shards is computed sequentially, and the migration of accounts is again under the control of a single blockchain node. However literature reports possible slow convergence speed for consensus algorithms and possible poor load-balancing solutions for the diffusion algorithm.

The convergence speed of diffusion and consensus algorithms has been thoroughly  analyzed in several publications. Convergence depends on several parameters, two of them, the diameter of the network and the initial load imbalance can be briefly discussed here.  The diameter of a network is the maximum distance between a pair of nodes, for example the diameter of a ring network is $\frac{n}{2}$, the diameter of a line network is $n-1$. In Table \ref{table 6} for a ring network of 5 nodes, the system converges to a state where the difference between the largest and smallest load is smaller/equal 1 after only 5 iterations. The same network topology with 10 nodes requires 17 iterations to achieve the same degree of load balance, 51 iterations with 20 nodes. Among regular graphs, ring networks have a relatively large diameter, it is well documented  that consensus algorithms embedded in such network have slow convergence. In regular graphs, the diameter depends on the degree of the nodes, thus sharded blockchains with large number of shards will have to be embedded in networks where shards are more highly interconnected.

The distribution of the initial loads has a lesser impact on the convergence behavior of the diffusion algorithm. Among the three shard blockchains that have been tested, Zilliqa has the largest initial load imbalance in absolute values (because of the differences in scale, 1 ZIL $\approx$ ETH0.00003182 according to \url{https://www.coingecko.com/en/coins/zilliqa/eth}). Our tests in Figures  \ref{f3} to \ref{f5} have been run using the absolute values in Figure \ref{f2} where the difference between the largest and the smallest load in Zilliqa is closed to 300 units, compared to 9 for Binance and 3 for Ethereum. Zelliqa required 36 iterations to reach a load balancing solution with a difference smaller than 1 among all the shards, while Binance required 12 iterations and Ethereum only 4 iterations. If we force the load imbalance to be smaller/equal to 0.03, which is the average cost of an Ethereum transaction, Ethereum needs 22 iterations and Binance 36. Considering much shorter epochs and 20 shards, literature \cite{WooSK0P20} reports initial load imbalances up to 65\% for sharded simulations of Ethereum. Using a similar initial Ethereum load imbalance for 10 shards, an unbalance $<$ 1 is achieved after 10 iterations and 52 iterations for an unbalance smaller/equal to 0.03.

There are two main reasons that may cause our implementation of the diffusion algorithm to return poor load-balancing solutions for sharded blockchains. One issue is algorithm \ref{phase1} can divide the load into infinitesimal values while account processing times are coarser, indivisible values. In some cases there may not exist a combination of the account processing times corresponding to the transfer vector values computed by algorithm \ref{phase1}. Thus the heuristic that computes subsets of accounts to migrate across shards may produce effective loads that differ with the values computed by the diffusion load-balancing algorithm. The second issue is related with the locality of the migration process in the diffusion algorithm. Each shard can only migrate its accounts  to its neighbors. Centralized algorithms such as LPT and MULTIFIT can schedule any account in the system to migrate to any shard. Thus centralized algorithms have a theoretical advantage. This is why in the paper we compare the load-balancing solutions based on the diffusion algorithm with solutions obtained using approximation algorithms such LPT and MULTIFIT. These comparisons ensure that the effective loads resulting from the diffusion algorithm  are within the approximation factors of the best centralized algorithms.

\section{Conclusion}
This paper proposes a consensus-based dynamic load-balancing algorithm for account-based sharded blockchains. The proposed algorithm is fully distributed, load-balance and accounts migration is computed and executed locally and periodically by each shard. Simulation tests of sharded blockchains have been run for three well-known blockchain platforms. Load-balancing solutions of the distributed algorithm were as good or better than two other centralized approximation algorithms that have approximation factors for solutions no further apart than 22\% above the optimal solution. 

Load-balancing in isolation may not by itself translate into overall improvement sharded blockchain performance. User account migration in some account-based permissionless sharded blockchain platforms incur fees which users may refuse to pay. Accounts migration may also increase the number of cross-shard transactions, increasing latency from transactions initiation to confirmation. In fact, load-balancing interacts or is bound by several design parameters of sharded blockchain. One possible future work direction will be to include new constraints in the current optimization model to represent some of these design parameters interacting with load-balancing in sharded blockchains. Such constraints will model for example increase latencies for cross-shard transactions or network communication costs for migrating accounts across shards.

\section*{Acknowledgement}
This work was funded by Gia Lam Urban Development and Investment Company Limited, Vingroup and supported by Vingroup Innovation Foundation (VINIF) under project code VINIF.2019.DA07
\bibliographystyle{splncs04}
\bibliography{citation}

\end{document}